\documentstyle[pre,aps,multicol,epsf,epsfig, amsmath, amssymb]{revtex}

\newcounter{saveapp}

\def\beginwide{
        \end{multicols} \vspace*{-0.5cm} \noindent
        \rule{3.5in}{.1mm}\rule{.1mm}{5mm} \widetext \medskip }
\def\beginwidetop{
        \end{multicols} \vspace*{-0.5cm} \noindent
        \widetext \medskip }
\def\endwide{
        \hspace*{3.35in}~\rule[-5mm]{.1mm}{5mm}\rule{3.5in}{.1mm}
        \begin{multicols}{2} \vspace*{-1.0cm} \noindent }
\def\endwidebottom{
        \begin{multicols}{2} \vspace*{-1.0cm} \noindent }

\def\beginwideapp{
	\setcounter{saveapp}{\value{section}}
        \end{multicols} \vspace*{-0.5cm} \noindent
        \rule{3.5in}{.1mm}\rule{.1mm}{5mm} \widetext \medskip 
	\appendix \setcounter{section}{\value{saveapp}}}
\def\beginwidetopapp{
	\setcounter{saveapp}{\value{section}}
        \end{multicols} \vspace*{-0.5cm} \noindent
        \widetext \medskip 
	\appendix \setcounter{section}{\value{saveapp}}}
\def\endwideapp{
	\setcounter{saveapp}{\value{section}}
        \hspace*{3.35in}~\rule[-5mm]{.1mm}{5mm}\rule{3.5in}{.1mm}
        \begin{multicols}{2} \vspace*{-1.0cm} \noindent  
	\appendix \setcounter{section}{\value{saveapp}}}
\def\endwidebottomapp{
	\setcounter{saveapp}{\value{section}}
        \begin{multicols}{2} \vspace*{-1.0cm} \noindent 
	\appendix \setcounter{section}{\value{saveapp}}}

\begin{document} 

\tightenlines

\title{Attractor neural networks storing multiple
space representations: a model for hippocampal place fields}
\author{F. P.  Battaglia and A. Treves}
\address{Neuroscience, SISSA, Via Beirut 2-4, 34014 Trieste -- Italy}

\maketitle

\begin{abstract}
 A recurrent neural network model storing multiple spatial maps, or
  ``charts'', is analyzed. A network of this type has been suggested as 
  a model for the origin of place cells in the hippocampus of rodents.
  The extremely diluted and fully connected limits are studied, and the
  storage capacity and the information capacity are found.
  The important parameters determining the performance of the network
  are the {\em sparsity} of the spatial representations and the degree of 
  connectivity, as found already for the storage of individual memory
  patterns in the general theory of auto-associative networks.

  Such results suggest a quantitative parallel between theories of
  hippocampal function in different animal species, such as primates
  (episodic memory) and rodents (memory for space).

\end{abstract}

\pacs{0000}

\begin{multicols}{2} 
\raggedcolumns 
 
 
\section{Introduction}

Knowledge of space is one of the main objects of computation by the brain.
It includes the organization of information of many kinds and many
origins (memory, the different sensory channels, and so on) into
mental constructs, i.e. {\em maps}, which retain a geometrical nature
and correspond to our perception  of the outer world.

Every animal species appears to have specialized systems for spatial knowledge
in some region of the brain, more or less well developed and capable of
performing sophisticated computations.
For rodents there is a large amount of experimental evidence that the
hippocampus, a brain region functionally situated at the end of all
the sensory streams, is involved in spatial processing.
Many hippocampal cells exhibit place related firing, that is, they
fire when the animal is in a given restricted region of the
environment (the ``place field''), so that
they contain a representation of the position of the animal in space.

The hippocampus, one of the most widely studied brain structures,
shares the same gross anatomical features across mammalian species,
nevertheless it is known to have different functional correlates for
example in primates and humans (where it is believed to be involved in
{\em episodic memory}, roughly, memory of events) and in rodents,
in which it is mainly associated  to spatial representation.

One relevant feature of the hippocampus which is maintained across
species is a region, named CA3, characterized by massive intrinsic
recurrent connections. It was  appealing for many theorists to
model this region as an {\em auto-associative memory} storing
information in its intrinsic synaptic structure, information which can
be retrieved from small cues by means of an attractor dynamics, and
which is represented in the
form of activity configurations.

Within the episodic memory framework, 
each attractor configuration of activity is the internal representation
of some memory item.
The CA3 auto-associative network can be seen as the heart of the
hippocampal system, containing the very complex, inter-modal
representations peculiar to episodic memory.
Auto-associative, or attractor neural networks have been extensively
studied with the tools of statistical physics \cite{amit:book}. 
Efficiency measures, like the number of storable memory items or the
quality of retrieval, can be computed for any appropriately defined
formal model.
An intensive effort was performed to embed in this idealized models
more and more elements of biological realism, trying to capture the
relevant anatomical and functional features affecting the performance
of the network.
It was then found that a theory of the hippocampus (or, more precisely,
CA3) as an auto-associative network has as its fundamental parameters
the degree of intrinsic connectivity, i.e. the average number of
units which send connections to a given unit, and the {\em sparsity}
of the representations, roughly the fraction of units which are active
in one representation. These parameters have biological correlates that are
measurable  with anatomical and neurophysiological techniques.

Spatial processing, as it is performed by the rodent hippocampus, also
involves memory of some kind; recent experimental evidence support the
idea that spatially related firing is not driven exclusively by sensory
inputs but also reflects some internal representation of the explored
environment.
First, place fields are present and stable also in the dark, and in
condition of deprived sensory experience. Second, completely different 
arrangements of place fields are found in  different environments
or even in the same environment in different behavioral conditions. 

These findings have led to hypothesize that CA3 (like, perhaps, other brain
regions with dense connectivity) stores ``charts'', representations
of  environments in the form of abstract manifolds, on which each
neuron corresponds to a point, that is the center of its place field.
Place fields arise as a result of an attractor dynamics, whose stable
states are ``activity peaks'' centered, in the chart space, at the
animal's position.
It is important to note that the localization of each neuron on a
chart does not appear to be related to its physical location in the
neural tissue.

The positions of place fields are encoded by the recurrent
connections, and it is possible to store many different charts in the
same synaptic structure, just as many different patterns are stored in
a Hopfield net, for example, and different activity peaks can be successively
evoked  by appropriate inputs, just as it happens with
auto-associative memories.

It is interesting to address the issue of whether an episodic memory network
and the ``spatial  multi-chart'' memory network share the same
functional contraints, so that a biological brain module capable
of performing one of the tasks is also adequate for the other.
Here we present a statistical mechanics analysis of the multi-chart
network, focusing on the parallel with autoassociative memory in the
usual (episodic memory) sense. It is found that the performances of these two networks
are governed by very similar laws, if the parallel between them is
drawn in the appropriate way.

In sec.~\ref{sec:singlemap} the case of a single attractor chart
stored is studied, then in sec.~\ref{sec:multimap} the case of
multiple stored charts is analyzed and the storage capacity is found,
first for a simplified model and then for a more complex model
which makes it possible to address the issue of sparsity of
representations.
In sec.~\ref{sec:info} the storable information in a multi-chart
network is calculated, making more precise the sense in which such a
network is a store of information, and completing the parallel
with auto-associative memories.

\section{The single map network} \label{sec:singlemap}

As a first step, we  consider the case of a single attractor map
encoded in the synaptic structure, as it was proposed in
\cite{tso:elba94}. We focus here on the shape and properties of the attractor states,
as a useful comparison for the following treatment of the multiple
charts case.

The neurons are modelled as threshold linear units, with
firing rate:
 
\begin{equation} 
V_i = g [ h_i - \theta]^+ = g (h_i - \theta) \Theta (h_i - \theta)
\end{equation}
i.e. equal to zero if the content of the square brackets is negative.
$h$ represents the synaptic input current, coming from other cells in
the same module, $\theta$ is a firing threshold, which may incorporate
the effect of a subtractive inhibitory input, common to all the cells,
as it will be illustrated later on.  The connectivity within the
module is shaped by the selectivity  of the units.  If
$\mathbf{r_i}$ is the position of the center of the place field of the
$i$-th cell, in a manifold $M$, of size $|M|$, corresponding to the environment, the 
connection between cells $i$ and $j$ may be expressed
as
\begin{equation}
J_{ij} = \frac{|M|}{N}  K(|{\mathbf r}_i - {\mathbf r}_j|), \label{Jij}
\end{equation}
where $K$ is a monotone decreasing function of its argument.

The synaptic input to the $i$-th cell is therefore given by

\begin{equation}
h_i = \sum_j J_{ij} V_j = \sum_j \frac{|M|}{N} K(|{ \mathbf r}_i - {\mathbf r}_j|) V_j.
\end{equation}

If the number $N$ of cells is large, and the place fields centers
(p.f.c.)  are homogeneously distributed over the environment 
$M$ (be it one or two-dimensional), we can replace the sum over the
index $j$ with an integration over the coordinates of the p.f.c:

\begin{equation}
h({\mathbf r}) = \int_M d{\mathbf r'} K(|{\mathbf r} - {\mathbf r'}|) V(r').
\label{hcontlim}
\end{equation}

Note that the normalization in eq.~\ref{Jij} is chosen in order to
keep the synaptic input to a given unit fixed when $|M|$ varies and
the number of units is kept fixed, that is, the density of p.f.c.s
$N/|M|$ varies (the $|M|$ factor will then compensate for the fewer
input units within the range of substantial $K$ strength).
A fixed-point activity configuration must have the form

\begin{equation}
V({\mathbf r}) = g\left[\int_M d{\mathbf r'} K(|{\mathbf r} -
{\mathbf r'}|) V({\mathbf r})' - \theta\right]^+. \label{eqV1}
\end{equation}

We could write eq.~\ref{eqV1} as

\begin{equation}
V({\mathbf r}) = \left\{
\begin{array}{lc}
 g(\int_{\Omega} d{\mathbf r'} K(|{\mathbf r} -
{\mathbf r'}|) V({\mathbf r'}) - \theta ) & {\mathbf r} \in \Omega 
\\  0 & {\mathbf r} \not\in \Omega 
\end{array} \right.
\end{equation} \label{1deqV}
where $\Omega$ is a domain for which there exists a solution of
eq.~\ref{1deqV} that is zero on the boundary.

If only solutions for which $\Omega$ is a convex domain are considered,
the fact that $V(r)$ is zero on $\partial \Omega$ will ensure that units
with p.f.c. outside $\Omega$ are under threshold, therefore their
activity is zero and solutions of eq.~\ref{1deqV} are guaranteed to be
solutions of eq.~\ref{eqV1}.  The size and the shape of the domain
$\Omega$ in which activity is different from zero is determined by
eq.~\ref{1deqV}.  As a first remark, we notice that it is
independent from the value of the threshold $\theta$. In fact, if
$V_\theta$ is a solution of (\ref{1deqV}) with threshold $\theta$,
given the linearity of eq.\ref{1deqV} within $\Omega$,
\[
V_{\theta'} = \frac{\theta'}{\theta} V_\theta
\]
will be a solution of the same equation with $\theta'$ instead of
$\theta$, with the same null boundary conditions on $\Omega$.  Rescaling
the threshold will then have the effect of rescaling the activity
configuration by the same coefficient.  This means that subtractive
inhibition cannot shape, e.g. shrink or enlarge, this stable
configuration, and therefore it is not relevant for a good part of the
subsequent analysis.  Some form of
inhibition is nevertheless necessary to prevent the activity from
exploding. Moreover, there are fluctuation modes which cannot be
controlled by overall inhibition as they leave the total average
activity constant. They will be treated in sec.~\ref{sec:stability}.
It is found that, at least in the one dimensional case, these modes do not affect
stability in the single chart case.

In absence of an external input, any solution can be at most
marginally stable, because a translation of the solution is again a
solution of eq.~\ref{1deqV}. An external, ``symmetry breaking'' input,
taken as small when compared to the contribution of recurrent synapses,
is therefore implicit in the following analysis.

\subsection{The one-dimensional case}

The case of a recurrent network whose attractors reflect the geometry
of a one-dimensional manifold, besides being a conceptual first step
in approaching the 2-dimensional case, is relevant by itself, for
example in modeling  other brain systems showing direction
selectivity, e.g. in head direction
cells \cite{taube:hdcrev96,redish:navig96}, and also for place fields on
one-dimensional environments \cite{tus:got97}.

In this case eq.~\ref{1deqV} reads:

\begin{eqnarray}
V(r) &=& g\left(\int_{-R}^{R} K(|r - r'|) V(r') - \theta\right) \nonumber
\\ V(R)&=& V(-R) = 0. \label{eqV1d}
\end{eqnarray}

For several specific forms of the kernel $K$ it is possible to solve
explicitly eq.~\ref{eqV1d}, yielding interesting conclusions.  For
example if:
\begin{equation}
K(|r - r'|) = e^{-|r - r'|}, \label{expker}
\end{equation}
(see also \cite{tso:elba94}) differentiating  eq.~\ref{eqV1d} twice  yields:
\begin{equation}
V''(r) = - \gamma^2 V(r) + g\theta,
\end{equation}
where $\gamma = \sqrt{2g-1}$.

Solutions vanishing at $-R$ and $R$ (and not vanishing in
$]-R, R[$), have the form:

\begin{equation}
V(r) = A \cos(\gamma r) + \frac{g\theta}{2g-1}, \label{solV1}
\end{equation}
with
\begin{equation}
A = -\frac{g\theta}{(2g-1) \cos(\gamma R)}.\label{solA1}
\end{equation}

The value of $R$ for which (\ref{solV1}) is a solution of
eq.~\ref{eqV1d} is determined by the integral equation itself: for
example, by
evaluating $V'(R)$ or $V'(-R)$ from eq.~\ref{eqV1d} we get:
\begin{equation}
V'(-R) = - V'(R) = g \theta. \label{cond-R}
\end{equation}

Substituting (\ref{solV1}) and (\ref{solA1}) in
eq.~\ref{cond-R} we have:
\[
\tan (\gamma R) = - \gamma
\]
so that
\[
R = \frac{\tan^{-1}(- \gamma) + n \pi}{\gamma}.
\]

Requiring $R$ to be positive and $V(x)$ to be positive for $-R < r
< R$, leads  to choose
\begin{equation}
R = \frac{-\tan^{-1}(\gamma) + \pi}{\gamma} \label{solR1},
\end{equation}
note that $A > 0$.

$R$ is then a monotone decreasing function of $\gamma$,
and therefore of the gain $g$.

This is also true for other forms of the connection kernel $K$. As an
example, consider the kernel:
\begin{equation}
K(r - r') = \cos(r - r')
\end{equation}
by a similar treatment it is shown that a solution is obtained with
\begin{equation}
R = \frac{1}{g}.
\end{equation}
The kernel
\begin{equation}
K(r - r') = \Theta(1 - | r - r' |) (1 - |r - r'|)
\end{equation} 
will result in a peak of activity of semi-width
\begin{equation}
R = \frac{\pi}{\sqrt{2  g}}.
\end{equation}

Equations of the type (\ref{eqV1}) have more solutions in addition
to the ones considered above, representing a single activity peak.
For example, if we  consider an infinite environment,
periodic solutions will be present as well, representing a row of
activity peaks separated by regions of zero activity.  These solution
can be checked to be unstable if we model inhibition as an
homogeneous term acting on all cells in the same way and depending on
the average activity. Intuitively, if we  perturb the solution by
infinitesimally displacing one of the peaks, it will tend to
collapse with the  neighbor which has come closer.

\subsection{The two-dimensional case}

To model  the place cells network in the hippocampus we need to
extend this result to a two dimensional environment.  The equation for
the neural activity will be:
\begin{equation}
V({\mathbf r}) = g \left[ \int_M d{\mathbf r'} K(|{\mathbf r} -
{\mathbf r'}|) V({\mathbf r'}) - \theta \right]^{+}. \label{2deq}
\end{equation}

The generalization to 2-D is straightforward if for the kernel
$K(|{\mathbf r} - {\mathbf r'}|)$ we consider the one with Fourier transform is
\begin{equation}
\hat{K}({\mathbf p}) = \frac{2}{1 + {\mathbf p}^2},
\end{equation}
(the two-dimensional analog of the kernel of
eq.~\ref{expker}) that is, a kernel resembling the propagator of a
Klein-Gordon field in Euclidean space.  The fact that this kernel
is divergent for $({\mathbf r} - {\mathbf r'}) \rightarrow 0$ does not
give rise to particular problems, since, in the continuum limit
of eq.~\ref{hcontlim}, the contribution to the field $h$
coming from the nearby points will stay finite, and in fact  two
units will be assigned p.f.c's so close to each other to yield a
overwhelmingly high connection only with a small probability.
Let us look for a solution with circular symmetry such that activity
$V({\mathbf r})$ is zero outside the circle of radius $R$,
${\cal C}(R)$.  If we apply the Laplacian operator on both sides of

\begin{equation}
V({\mathbf r}) = g \int_{{\cal C}(R)} d{\mathbf r'} K({\mathbf r} -
{\mathbf r'}) V({\mathbf r'}) - \theta \label{eqcirc2d}
\end{equation}
we obtain:
\begin{equation}
\nabla^2 V({\mathbf r'}) = - \gamma^2 V({\mathbf r'}) + g\theta
\end{equation}
(again, $\gamma^2 = 2g-1$), which in polar coordinates reads:

\begin{equation}
V''(r) + \frac{1}{r} V'(r) = - \gamma^2 V(r) + g\theta.
\end{equation}	 

The solution is
\begin{equation}
V(r) = A J_0 (\gamma r ) + \frac{g\theta}{2g - 1}. \label{sol2d}
\end{equation}

$J_0$ is the Bessel function of order 0. For the solution to vanish on the boundary of ${\cal C}(R)$
one must take:

\[
A = \frac{g\theta}{(2g - 1)J_0 (\gamma R )}.
\]

The other condition that	
determines $R$ may be found by substituting (\ref{sol2d}) in
eq.~\ref{eqcirc2d}. Here again, $R(g)$ is a monotone decreasing function.

As in the one-dimensional case, solutions with a non-connected (or
even non-convex) support can be seen not to be stable.

\section{Storing more than one map} \label{sec:multimap}

Let us imagine now that the p.f.c's for each cell are drawn with
uniform distribution on the environment manifold $M$, and connections
are formed according to (\ref{Jij}).  Several ``space
representations'' may be created by drawing again at random the r.p.c. of
each cell from the same distribution.  The connection between each
pair of cells will then be  the sum of a number of terms of the
form (\ref{Jij}), one for every ``space representation'', or
``map'', or ``chart''. With $p= \alpha N$ maps, and the r.p.c. of the $i$-th cell in
the $\mu$-th map indicated by ${\mathbf x}_i^{(\mu)}$:
\begin{equation}
J_{ij} = \sum_{\mu = 1}^{p} \frac{|M|}{N}K(|{\mathbf r}_i^{(\mu)} -
{\mathbf r}_j^{(\mu)}|). \label{Jijmulti}
\end{equation}

The question that immediately arises is: what is the capacity of this
network, that is, how many maps can we  store, so that stable activity
configurations, corresponding to some region in the environment described by
one map, like the ones described by the solutions of eq.~\ref{1deqV},
are present?  The problem resembles the classic attractor neural
network problem \cite{amit:book}, with threshold linear units.  A
standard treatment has been developed \cite{treves:cap90} allowing to
calculate the capacity of a network of threshold linear units with
patterns drawn from a given distribution and stored by means of 
a hebbian rule.  The treatment is very simplified in the extreme
dilution limit \cite{derrida:dilu87,treves:dilu91}. In the next sections it will be
shown how this treatment can be extended to the map case, first for
one particular form of the kernel $K$, leading to the solution of the
capacity problem for a fully connected network; 
in the following, the solution is extended to more general kernels, first in the diluted
limit, then for the fully connected network.

Another related question is: how much information is the synaptic
recurrent structure encoding, and in which sense is the synaptic 
structure  a store of information? The aim is to develop a full
parallel between the multi-chart network and autoassociative networks,
and if possible to characterize the  parameters
constraining the performance of this system.

\subsection{The fully connected network: ``dot product'' kernel}

Let us consider a manifold $M$ with periodic boundary
conditions, that is, a circle in one dimension and a torus in two
dimensions.  The p.f.c. of a cell ${\mathbf r}_i$ can then be described
by a 2-dimensional unit vector $\vec{\eta}_i$ for the one-dimensional
case and by a pair of unit vectors $\vec{\eta}_i^{1,2}$ for the two
dimensional case.  Suppose now that the contribution from the $\mu$-th
map to the connection between cell $i$ and cell $j$ is given by:
\begin{equation}
K(|{\mathbf r}_i^{(\mu)} - {\mathbf r}_j^{(\mu)}|) = \sum_{l=1}^{d}
(\vec{\eta}_i^{l(\mu)} \cdot \vec{\eta}_j^{l(\mu)} +1 ) ,
\label{Kdotprod}
\end{equation}
so that
\begin{equation}
J_{ij} = \frac{1}{N} \sum_{\mu = 1}^{p} \sum_{l=1}^{d} (\vec{\eta}_i^{l(\mu)}
\cdot \vec{\eta}_j^{l(\mu)}+1) , \label{Jijdotprod}
\end{equation}
where $d$ is the dimensionality.

$p = \alpha N$ is the number of stored charts. 
Eq.~\ref{Kdotprod} describes an excitatory, very wide spread form for
the kernel (\ref{Jij}) (the
contribution to the connectivity is zero only if the r.p.c.s of the two
cells are at the farthest points apart, i.e. at $180^{\circ}$). This spread of connectivity would lead to configurations
of activity that are large in the r.p.c. space, that translated in 
auto-associative memory language would be very ``unsparse'', i.e. very
distributed
representations.  It is therefore plausible that this will severely
limit the capacity of the net.  In any case, the form of
(\ref{Kdotprod}), factorizable in one term depending on
$\vec{\eta}_i$ and one term depending on $\vec{\eta}_j$, after
incorporating the constant part in a function $b^0(x)$, makes it
possible to perform the free energy calculation through Gaussian
transformations as in \cite{treves:cap90}.
A similar model has been studied in \cite{tus:sam97} with McCulloch-Pitts
neurons.

A Hamiltonian useful to describe the thermodynamics of such a system is

\begin{eqnarray}
H =&& - \frac{1}{2} \sum_{i,j(\neq i)} J_{ij} V_i V_j - N B \left(
\sum_i \frac{V_i}{N} \right) - \nonumber \\
&& \sum_l \sum_i \sum_{\mu}
{\mathbf s}^{l(\mu)} \cdot \vec{\eta}^{l(\mu)}_i V_i \label{dotprod:ham}
\end{eqnarray}
where $B(x) = \int^x b(y) dy$, and $b(x)$ is a function describing
an uniform inhibition term depending on the average activity in the
net.  ${\mathbf s}^{l(\mu)}$ is a symmetry breaking field, pointing in
a direction in the $\mu-th$ map space.  The mean field free-energy in
the replica symmetric approximation can be calculated (the partition
function is calculated as the trace over a measure that implements the
threshold-linear transfer function, see \cite{treves:cap90}).  The
presence of a phase with spatially specific activity correlated with
one map will be signaled by solutions of the mean field equations
with a non-zero value for the order parameter
\begin{equation}
{\mathbf x}^{l(\mu)} = \frac{1}{N d} \sum_{i=1}^{N}
\vec{\eta}^{l(\mu)}_i V_i
\end{equation}
which plays the role of the overlap in an auto-associative memory.
This parameter has the meaning of a population vector \cite{geopopvec},
that is, the animal position is indicated by an average over p.f.c.s
of the cells weighted by cells activity.

The set of resulting mean field equations can be reduced to a set of two equations, eqs.~\ref{eqE1} and \ref{eqE2}, in two variables, the ``non-specific'' signal-to-noise ratio, $w$, and the ``specific'', space related signal-to-noise ratio $v$.
The details of the calculation are reported in Appendix \ref{app:dotprod}.

The critical value  $\alpha_c$ indicating the  storage capacity
of the network is the maximum value for which eq.~\ref{eqE1} still
admits solutions corresponding to space related activity (non-zero $v$) and may
be found numerically. 
At this value $\alpha_c$ the system undergoes a first-order
phase-transition towards a state in which no space-related activity is
possible.
Eq.~\ref{eqE2} gives the range of gain values for which there exist solutions at a given $\alpha \leq \alpha_c$ \cite{treves:cap90}.

In this model there is no possibility for modulating the spread of
connections in the chart-space. As we anticipated,  the activity configurations that
one obtains are very wide, with a large fraction of units active at
the same time. Cells will have very large place fields, covering a
large part of the environment (of the order of roughly one half for
the one dimensional case, and roughly one quarter for the two-dimensional
case).
As one would infer from the analysis of  autoassociative memories
storing patterns, for example binary, these ``unsparse''
representations of space will lead to a very small capacity of the
net.

For  the model defined on the one-dimensional circle the 
capacity  found is $\alpha_c \sim 0.03$. At this value the system undergoes a
first order transition. As $\alpha$ increases beyond $\alpha_c$,
${\mathbf x}$ jumps discontinuously from a finite value to zero.

The capacity for the diluted analogue of this model (see
\cite{treves:dilu91}, Appendix A and section~\ref{sec:diluted}) is given by the equation
\begin{equation}
E_1(w,{\mathbf v}) \equiv [(1 + \delta) A_2]^2 - \alpha A_3 = 0.
\label{eqE1dilu}
\end{equation}
Remember that in this case $p = \alpha c N$ where $c$ is the 
connectivity fraction parameter, see section~\ref{sec:diluted}.
In this case $\alpha_c \sim 0.25$. At $\alpha_c$ the transition 
is second order, with the
``spatial overlap'' ${\mathbf x}$ approaching continuously zero,
verified at
least with  the precision at which it was possible to to solve
numerically eq.~\ref{eqE1dilu}.
For the 2-D case,  storage capacities are $\alpha_c
\sim 0.0008$ for the fully connected network and $\alpha \sim 0.44$ for
the diluted network. 

To get a larger capacity, and to provide a possible
comparison with the
experimental data from the hippocampus, in which the tuning of place
fields is generally narrow, we must  extend our treatment to
more general kernels, and this will be done in the following two
sections.

\subsection{Generic kernel: extremely diluted limit} \label{sec:diluted}

Consider a network in which every threshold-linear unit, whose
activity is denoted by $V_j$, senses a field:
\begin{equation}
h_i = \frac{1}{c} \sum_{j \neq i} C_{ij} J_{ij} V_j,
\end{equation}
where $J_{ij}$ is given by eq.~\ref{Jijmulti}. From now on the kernel
$K$ is defined as 
\begin{eqnarray}
K(\vec{r} - \vec{r}') &=& \hat{K}(\vec{r} - \vec{r}') -
\bar{K} 
\nonumber\\
\bar{K} &=& \left\langle \left\langle \hat{K}(\vec{r} -
\vec{r}') 
\right\rangle
\right\rangle  \label{dilu:condK}
\end{eqnarray}
for any $\vec{r}$,
where  $\langle \langle \ldots \rangle \rangle$ means averaging over
$\vec{r}$.
With this notation, whatever the original kernel $\hat{K}$, $K$ is the
subtracted kernel which averages to zero.
The manifold $M$ is taken with periodic boundary condition
(that is a circle in one dimension and a torus in the two dimensional
case).

$C_{ij}$ is a
``dilution matrix'' 
\begin{equation}
C_{ij} = \left\{ 
	\begin{array}{lc}
	1 & \text{with prob. $c$,} \\
	0 & \text{with prob. $1-c$}
	\end{array} \right.
\end{equation} 
and 
$Nc/ \log N \rightarrow 0$ as $N \rightarrow \infty$.
In the thermodynamic limit $N \rightarrow \infty$ the activity of any two
neurons $V_i$ and $V_j$ will be uncorrelated \cite{derrida:dilu87}.
A number of charts $p = \alpha c N$ is stored.
Looking for solutions with one ``condensed'' map, that is,  solutions
in which activity is confined to units having p.f.c. for a given chart
in a certain neighborhood, it is possible to write the field $h_i$ as
the sum of two contributions, a ``signal'', due to the condensed 
map and a ``noise'' term, $\rho z$ -- $z$ being  a random variable with
Gaussian distribution and variance one --  due to  all the other, uncondensed, maps,
namely, in the continuum limit, labeling units with the position
$\vec{r}^1$ of their p.f.c. in the condensed map,
\begin{equation}
h(\vec{r}^1) = g \int_M d\vec{r}^{1'} K(\vec{r}^1 -
\vec{r}^{1'})V(\vec{r}^{1'}  ) + \rho z; 
\end{equation}
the noise will have a variance
\begin{equation}
\rho^2 = \alpha y |M|^2 \langle \langle  K^2(\vec{r} -
\vec{r}')  \rangle \rangle, \label{dilu:eqrho}
\end{equation}
where
\begin{equation}
y = \frac{1}{N} \sum_{i=1}^{N} \langle V_{i}^{2} \rangle.
\end{equation}

The fixed point equation for the average activity profile
$x^1(\vec{r})$
is 

\begin{equation}
x^1(\vec{r}) = g \int^+ Dz (h(\vec{r}) - \theta).
\end{equation}
where again $Dz$ is the Gaussian measure, and 
\begin{equation}
h(\vec{r}) = \int d \vec{r}' K(\vec{r} -
\vec{r}')  x^1(\vec{r}') + b(x) - \rho z 
\end{equation} 
and 
\begin{equation}
x =  \int \frac{d \vec{r}}{|M|} \; x^1(\vec{r} ) 
\end{equation}
is the average overall activity.
The average squared activity (entering the  noise term) will read:
\begin{equation}
y = g^2  \int \frac{d \vec{r}}{|M|} \int^+ Dz (h(\vec{r}) - \theta)^2.
\end{equation}

The fixed point equations may be solved  introducing the rescaled
variables
\begin{eqnarray}
w = \frac{b(x) - \theta}{\rho} \\
v(\vec{r}) = \frac{x^1(\vec{r})}{\rho}.
\end{eqnarray}

The fixed point equation for $v(\vec{r})$ is
\begin{equation}
v(\vec{r}) = g {\cal N}\left( \int d \vec{r}' K(\vec{r} -
\vec{r}') v(\vec{r}') + w \right) \label{dilu:eqv}
\end{equation} 
where
\begin{equation}
{\cal N}(x) = x \Phi(x) + \sigma(x),
\end{equation}
($\Phi(x)$ and $\sigma(x)$ are defined in eq.~\ref{eqPhi} and
eq.~\ref{eqsigma})  is a ``smeared threshold linear function'',
monotonically  increasing, with 
\[
\lim_{x \rightarrow -\infty} {\cal N} (x) = 0
\]
and
\[
 \lim_{x \rightarrow +\infty} {\cal N} (x) / x = 1.
\]

In terms of $w$  and $v(\vec{r})$,  $y$ reads:
\begin{equation}
y = \rho^2 g^2  \int  \frac{d \vec{r}}{|M|} {\cal M} \left( \int d \vec{r}' K(\vec{r} -
\vec{r}') v(\vec{r}') + w \right) \label{dilu:eqy}
\end{equation}
where
\begin{equation}
{\cal M}(x) = (1+x^2)\Phi(x) + x \sigma(x).
\end{equation}

By substituting eq.~\ref{dilu:eqy} in eq.~\ref{dilu:eqrho},
we obtain
\begin{equation}
\frac{1}{\alpha} =  g^2 |M| \langle \langle K \rangle \rangle \int d \vec{r} {\cal M} \left( \int d \vec{r}' K(\vec{r} -
\vec{r}') v(\vec{r}') + w \right). \label{dilu:eqalpha}
\end{equation} 

If we can solve eq.~\ref{dilu:eqv} and find $v(\vec{r})$ as a
function of $w$ and $g$, a solution is found corresponding to a value
of $\alpha$ given by eq.~\ref{dilu:eqalpha}.
To find the critical value of $\alpha$, we have to maximize $\alpha$
over $w$ and $g$.

The mathematical solution of eq.~\ref{dilu:eqv} is treated in Appendix
\ref{app:gendilu}.

\begin{figure}[th]
\narrowtext
\begin{center}
\epsfig{file=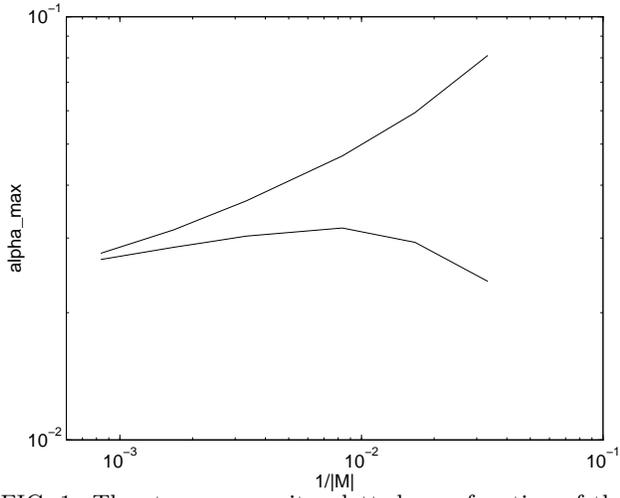, width=\columnwidth}
\caption[123]{The storage capacity plotted as a function of the
``map sparsity'' $a_m$, for the 1-D model, for the extremely
diluted (upper curve) and the fully connected (lower curve) limit.} 
\label{fig:alpha1}
\end{center}
\end{figure}

\begin{figure}[th]
\narrowtext
\begin{center}
\epsfig{file=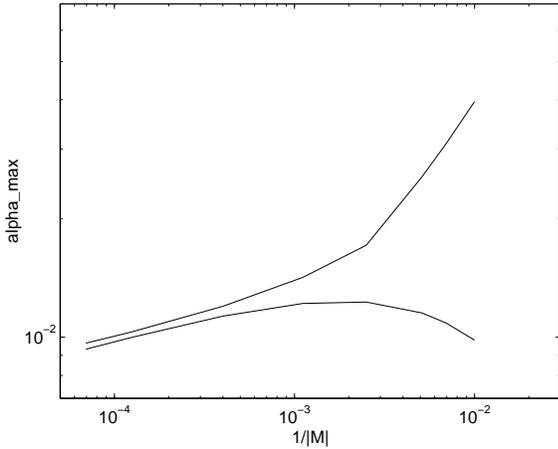, width=\columnwidth}
\caption{Same as fig.~\ref{fig:alpha1}, for the 2-D model. The
capacity is smaller than for the 1-D model for the same $a_m$.} 
\label{fig:alpha2}
\end{center}
\end{figure}

\begin{figure}[th]
\narrowtext
\begin{center}
\epsfig{file=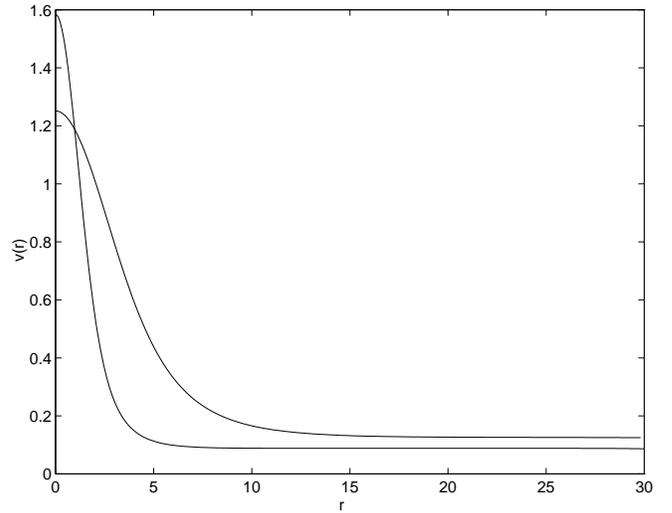, width=\columnwidth}
\caption[123]{The ``activity peak'' profile corresponding to the
solution of eq.~\ref{dilu:eqv} at the maximal storage level at $|M| =
30$ and $|M| = 15$. The second case is plotted expanded to match the
environment size of the first one and to show the effect of more widespread
connections.} 
\label{fig:sol1}
\end{center}
\end{figure}

\begin{figure}[th]
\narrowtext
\begin{center}
\epsfig{file=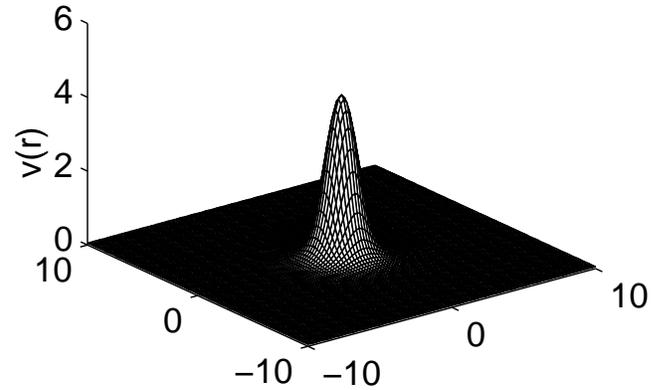, width=\columnwidth}
\caption[123]{The maximal storage activity peak profile in 2-D at $|M| = 400$.
} 
\label{fig:sol2}
\end{center}
\end{figure}

With this model, we can modulate the spread of connections by
acting on $K(\vec{r} - \vec{r}')$  or alternatively, by varying the size of
the environment.  The results are depicted in fig.~\ref{fig:alpha1}  for the 1-D
circular environment and in fig.~\ref{fig:alpha2} for the 2-D toroidal
environment (upper curves). Examples of the solutions of
eq.~\ref{dilu:eqv} are displayed in fig.~\ref{fig:sol1} for the 1-D
environment and in fig.~\ref{fig:sol2} for the 2-D environment.

We note that, as the environment gets larger in comparison to the
spread of connections (therefore, to the size of the
activity peak), the capacity decreases approximately as 
\begin{equation}
\alpha_c \sim - 1/ \log(a_m) \label{formalphac}
\end{equation}
where $a_m$ is the {\em map sparsity} and it is equal to:

\begin{equation}
a_m = \frac{k_d}{|M|}
\end{equation}
where $k_d$ is a factor roughly equal to $\sim 4.5$ for the 1-D model
and $\sim 3.6$ for the 2-D model.

That is, the sparser the coding, the less the capacity.
This is, at first glance, in contrast with what is known from the
theory of auto-associative networks, in which sparser representations
usually lead to larger storage capacities.

For comparison, keeping the formalism of \cite{treves:cap90}, for  threshold-linear
networks with hebbian learning rule, encoding memory patterns $\{ r_i \}_{i =
 1 \dots N}$ with sparsity $a$ defined as
\[
a_p = \frac{ \langle\langle r \rangle\rangle^2}{ \langle\langle
r^2 \rangle\rangle}
\]
(for binary patterns this is equal to the fraction of active units), and for
small $a$, the capacity is given by

\begin{equation}
\alpha_p \sim \frac{1}{a_p \log(1/a_p)}. \label{formalphap}
\end{equation}

The apparent paradox 
(larger capacity with sparser patterns, smaller with sparser charts)
is solved as one recognizes that each chart can be seen as
a collection of configurations of activity relative to different points 
in space covering, as in a tiling, the whole environment.
Each configuration is roughly equivalent to a pattern in the usual
sense. 
Intuitively, and in sense that will made clearer below, a chart is
equivalent, in terms of ``use of synaptic resources'' to a
number proportional to $a_m^{-1}$ of patterns of sparsity $a_m$.

The proportionality coefficient, or equivalently, the distance at which
different configurations are to be considered to establish a correct
analogy, will be dealt with in Appendix~\ref{app:info}.

These considerations and the comparison of eq.~\ref{formalphac} and
eq.~\ref{formalphap} make clear that $\alpha_c$ is the exact analogue
of the pattern autoassociators' $\alpha_p$.

{\bf Figures 1, 2, 3, 4, approx. here}

\subsection{Inhibition independent stability} \label{sec:stability}

The dynamical stability of the solutions of eq.~\ref{dilu:eqv} is in
general determined 
by the precise functional form chosen for the inhibition, which we
assumed a function of the average overall activity in the net.
Nevertheless, there are fluctuations modes which leave the average
activity unaltered. Stability against these modes is therefore
unaffected by the inhibition and may be checked already for a general model.
Let us consider a ``synchronous'' dynamics, that is, all the neurons
are updated simultaneously at each time step. 
The evolution operator for the variables $V(r, t)$ and $\rho(t)$
is:

\beginwide
\begin{eqnarray}
V(r, t+1) &=& g \rho(t) {\cal N} \left( \int_M \frac{d r'}{|M|}
K(r - r') \frac{V(r', t)}{\rho(t)} + \frac{b(x(t))}{\rho(t)}
\right) \label{stab:evolV} \\
\rho^2(t+1) &=& g^2 \alpha |M| \rho^2(t) \langle \langle K^2 \rangle \rangle
\int_M dr {\cal M} \left( \int_M \frac{d r'}{|M|}
K(r - r') \frac{V(r', t)}{\rho(t)} + \frac{b(x(t))}{\rho(t)}
\right). \label{stab:evolrho}
\end{eqnarray}

This evolution operator has as its fixed points $V_0(r) = \rho_0
v_0(r)$ and $\rho_0$ where $v_0(r)$ and $\rho_0$ are the solutions of
eq.~\ref{dilu:eqv}, \ref{dilu:eqrho}, and ~\ref{dilu:eqy}, i.e. the
stable states of our system.

We can linearize the evolution operator around $(V_0(r), \rho_0)$
and look for fluctuation modes (eigenvectors) $(\delta V(r),
\delta \rho)$ with
\begin{equation}
\int_M dr \, \delta V(r) = 0 \label{stab:constavV}
\end{equation}

We obtain the following equations:
\begin{eqnarray}
\lambda \delta V(r) &=& g \Phi(u_0(r)) \left[ \int_M dr'
K(r - r') \delta V(r') \right] + g \sigma(u_0(r)) \delta
\rho  \label{stab:eigenV}\\
\lambda \delta \rho &=& \left(1 - \frac{1}{2} g \alpha |M| \langle \langle
K^2 \rangle \rangle  \int_M d r u_0(r) v_0(r) \right) \delta \rho + \frac{1}{2} g \alpha |M| \langle \langle
K^2 \rangle \rangle \int_M dr u_0(r) \delta V(r) 
\label{stab:eigenrho},
\end{eqnarray}
where
\[
u_0(r) = {\cal N}^{-1} \left(\frac{v_0(r)}{g}\right).
\]

Inserting eq.~\ref{stab:eigenV} in eq.~\ref{stab:constavV}:
\begin{equation}
\delta \rho = - \frac{ \int_Md r' \Phi(u_0(r)) \left[ \int_M dr
K(r - r') \delta V(r') \right]}
{\int_M dr  \sigma(u_0(r))} \label{stab:rhoconstV}
\end{equation}

\endwide

Eq.~\ref{stab:rhoconstV} can be inserted again in eq.~\ref{stab:eigenV},
obtaining a closed integral equation in $\delta V$.
Unfortunately, this equation is very difficult to solve,
but we can derive a stability condition by making an ansatz on the
form of the eigenfunction $\delta V(r)$. More precisely, let us
concentrate on the 1-D case. We look for solutions with even symmetry
(we know there must be an eigenfunction with odd symmetry, and
eigenvalue equal to 1, corresponding to a coherent displacement of the
activity peak). This kind of solution corresponds to spreading and
shrinking of the activity peak. Let us assume that the even
eigenfunction with the highest eigenvalue (the most unstable) has
only  two nodes (an even eigenfunction must have at least two
nodes because of eq.~\ref{stab:constavV}), at $r_0$ and $-r_0$.
Let us take the sign of the eigenfunction $\delta V(r)$ such that
$\delta V(0) > 0$.
From eqs.~\ref{stab:constavV} and \ref{stab:rhoconstV} we see
that 
\[
\delta \rho < 0.
\]
Now, from eq.~\ref{stab:eigenrho}: 

\begin{eqnarray}
\lambda  =&& \left(1 - \frac{1}{2} g \alpha |M| \langle \langle
K^2 \rangle \rangle  \int_M d r u_0(r) v_0(r)\right)  +
\nonumber \\
&&\frac{1}{2} g \alpha |M| \langle \langle
K^2 \rangle \rangle \int_M dr u_0(r) \frac{\delta V(r)
}{\delta \rho},
\end{eqnarray}
and we recognize that 
\[
\int_M dr u_0(r) \frac{\delta V(r)
}{\delta \rho} < 0.
\]

Thus,
\begin{equation}
\lambda < 1 - \frac{\Gamma}{2}
\end{equation}
with 
\begin{equation}
\Gamma =  g \alpha |M| \langle \langle
K^2 \rangle \rangle  \int_M d r u_0(r) v_0(r).
\end{equation}

Thus, if the ansatz we formulated holds, we have a stability
condition $\Gamma > 0$, which is found to be fulfilled for all the solutions we
found relative to maximal storage capacity. This implies that the
storage capacity result is not affected by instability of the
solutions, provided of course that an appropriate form for inhibition is
chosen.
This stability result is also related to the correlation in the
static noise for two solutions centered at different p.f.c.s, as we will show
in App.~\ref{app:info}.

It can  also be shown that by taking the $\alpha \rightarrow 0$ limit
(i.e. the single
chart case), one always has $\Gamma > 0$ since
it is $v_0(r) = 0$ when $u_0(r) < 0$.
\subsection {The fully connected model} \label{sec:fullyconn}

The treatment of the model with the fully connected network and
a kernel $K$ for connection weights satisfying the condition
(\ref{dilu:condK}) will use the replica trick to average over the
disorder (the realizations of the $\vec{r}$'s) and will
eventually lead to a non-linear integral equation for the average
activity profile in the space of the ``condensed map'' very similar to
eq.~\ref{dilu:eqv}.
Let the Hamiltonian of the system be
\begin{eqnarray}
H = && - \frac{1}{2} \sum_{i,j(\neq i)} J^c_{ij} V_i V_j - N B \left(
\sum_i \frac{V_i}{N} \right) - \nonumber \\
&& \sum_i \sum_{\mu}
{\mathbf s}^{l(\mu)} \cdot r^{(\mu)}_i V_i \label{fc:ham}
\end{eqnarray} 
where now the $J^c_{ij}$ are given by (\ref{Jijmulti}) with a generic
kernel  
\begin{equation}
K(\vec{r} - \vec{r}') = \hat{K}(\vec{r} -
\vec{r}') - \bar{K}
\end{equation}
where, again, 
\[
\bar{K} = \langle \langle \hat{K}(\vec{r} - \vec{r}' \rangle \rangle.
\]

The free energy calculation is sketched in Appendix
\ref{app:genfull}. 
Again, the stable states of the system are governed by mean field
equations.
The mean field equation, eq.~\ref{fc:eqv} is an integral equation in the
functional order parameter $v(\vec{r})$, the average space profile
of activity.

If we are able to solve eq.\ref{fc:eqv} and find
$v^\sigma(\vec{r})$ as a function of $w$ and  $g'$, by
substituting eq.~\ref{fc:eqy} and eq.~\ref{fc:eqpsi} in
eq.~\ref{fc:eqrho} we have an
equation that gives us the value of $\alpha$ corresponding to that
pair $(g',w)$.
$\alpha_c$ is then the maximum of $\alpha$ over the possible values of
$(g',w)$.

To solve eq.~\ref{fc:eqv}, it is easy to verify that if
$\tilde{v}(\vec{r})$ is a solution of 
\begin{equation}
\tilde{v}(\vec{r}) = g' {\cal N}\left( \int_M d \vec{r}' \hat{K}(\vec{r} -
\vec{r}') \tilde{v}(\vec{r}') + \hat{w} \right)  \label{fc:eqtilde}
\end{equation}
with 
\[
\hat{w} = w - \bar{K} \int_M d \vec{r} \; \tilde{v}(\vec{r})
\]
that is, the same equations as eqs.~\ref{dilu:eqhat} and
\ref{dilu:eqw}, then
\[
v(\vec{r}) = \int_M d \vec{r}' \left[
L(\vec{r} - \vec{r}') - \bar{L} \right]
\tilde{v}(\vec{r}') 
\]
is a solution of eq.~\ref{fc:eqv}. $\tilde{v}$ can therefore be
interpreted as the average activity profile, apart from a constant.
Eq.~\ref{fc:eqtilde} can be solved with the same procedure used for
eq.~\ref{dilu:eqv}, and the maximum value of $\alpha$ can be found by
maximizing over $g'$ and $\hat{w}$.

The results for 1-D and 2-D environment are
depicted in fig.~\ref{fig:alpha1} and  \ref{fig:sol2} (lower curves).
As we may expect from pattern autoassociator theory, the  capacity is
much lower than for the diluted model, due to an increased 
interference between different charts..
As the sparsity $a \sim 1 / |M|$ gets smaller, the capacities of the two models
get closer, both being proportional to $\frac{1}{\log(|M|)}$.
Reducing the sparsity parameter of space representations has
therefore the effect of minimizing the difference between nets with
sparse and full connectivity.

\subsection{Sparser maps}

A possible extension of this treatment is inspired from the
experimental finding that, in general, not all the cells have place
cells in a given environment. 
Ref.~\cite{wilmn93} e.g. reported that $\sim 28-45  \%$ of pyramidal cells of CA1
have a place field in a certain environment.
We would like to see how this fact could affect the performance of the
multi-chart auto-associator.
It is then natural to introduce a new sparsity parameter, the {\em
chart sparsity} $a_c$ indicating the fraction of cells which
participate in a chart.
We will show that, for the capacity calculation, $a_c^{-1}$
``sparse'' charts are equivalent to a single ``full'' chart, of size
$a_c^{-1}$.
We will present the argument for the diluted case, the fully connected
case is completely analogous.

Let $m_i^\mu$ be equal to $1$ if cell $i$ participates in chart $\mu$,
that is with probability $a_c$.
Thus, the synaptic coupling $J_{ij}$ will read 
\begin{equation}
J_{ij} = \sum_{\mu = 1}^{p} \frac{|M|}{a_c N} K({\mathbf x}_i^{(\mu)} -
{\mathbf x}_j^{(\mu)}) m^\mu_i m^\mu_j. \label{Jijmultisparse}
\end{equation}

Let us consider a solution with one condensed map: cells participating
with p.f.c. in $r$ in
that map will have a space related signal-to-noise ratio 
\begin{equation}
v(\vec{r}) = g {\cal N}\left( \int d \vec{r}' K(\vec{r} -
\vec{r}') v(\vec{r}') + w \right) \label{dilu:eqvsparse}
\end{equation}

for all the neurons {\em not} participating in the condensed map we
will have 
\begin{equation}
v = {\cal N}(w).
\end{equation}
The noise will have a variance
\begin{equation}
\rho^2 = \alpha a_c y \left(\frac{|M|}{a_c}\right)^2 \langle \langle  K^2(\vec{r} -
\vec{r}^\mu  \rangle \rangle, \label{dilu:eqrhosparse}
\end{equation}
that is $a_c$ times the value we would get for the same number of ``full'' charts with
size $|M|/a_c$.
and now 
\begin{eqnarray}
y = &&\rho^2 g^2 \left\{a_c\int_M \frac{d \vec{r}}{|M| } {\cal M} \left( \int d \vec{r}' K(\vec{r} -
\vec{r}') v(\vec{r}') + w \right) \right. + \nonumber \\
&& \left.(1 - a_c) {\cal M}(w) \right\}. \label{dilu:eqysparse}
\end{eqnarray}

By comparing eq.~\ref{dilu:eqy} and eq.~\ref{dilu:eqysparse}, and
remembering that for $\vec{r}$ far from the activity peak,
$v(\vec{r}) \sim {\cal N}(w)$  we
realize that this $y$ value is approximately equivalent to the $y$ value  we would get for ``full''
charts of size $|M|/a_c$.

Inserting eq.~\ref{dilu:eqrhosparse} and eq.~\ref{dilu:eqysparse} in
eq.~\ref{dilu:eqalpha}, one finds, for the maximal capacity:
\begin{equation}
\alpha_{c\, \text{(sparse charts)}} \sim \frac{1}{a_c \log(\frac{|M|}{a_c})}. \label{dilu:eqalphasparse}
\end{equation}

As we anticipated one may interpret this result as follows: the capacity
is the same as if we had taken $a_c^{-1}$ ``sparse'' charts, including
$\sim N$  cells, and put them side by side to form one single ``full''
chart.
If we have started with $\alpha  C$ ``sparse'' charts we now have have
$a_c \alpha  C$ ``full  charts''.
From eq.~\ref{dilu:eqalpha} we see that we can store at most 
$\alpha_{c \,\text{(full charts)}}  C$ full charts and
\[
\alpha_{c\, \text{(full charts)}} \sim \frac{1}{ \log(\frac{|M|}{a_c})}
\]
and this explains eq.~\ref{dilu:eqalphasparse}.
Therefore, this network is {\em as efficient} in terms of
spatial information storage as the one operating with full charts.

\section{information storage} \label{sec:info}

Like a pattern auto-associator, the chart auto-associator is an
information storing network.
The cognitive role of such a module could be to provide a
spatial context to information of non-spatial nature contained in other
modules, which connect with the multi-chart
module.
Each chart represents a different spatial organization, possibly
related to a different environmental/behavioral condition.
Within each chart, a cell is bound to a particular position in space,
thus being the means for attaching some piece of knowledge to a particular point
in space, through inter-module connections.
To give a very extreme, unrealistic, but perhaps useful,  example, let
us assume that each cell encodes a particular discrete item, or the memory of some events happened somewhere in the
environment, as in a ``grandmother cell'' fashion, encoding ``the grandmother
sitting in the armchair in the dining room''. The encoding of the
``grandmother''  may be
accomplished by some set of afferents from other modules.
The multi-chart associator can then attach a spatial location to that
memory of the ``grandmother''.
The spatial location encoded is ideally represented for each cell by its p.f.c.

In this sense, the information encoded in the network, which can be
extracted by measures of the activity of the units, is the information
about the spatial tuning of the units, that is their p.f.c.s.

To restate this concept in a formal way, we look for
\begin{equation}
I_s = \lim_{S \rightarrow \infty}  \frac{1}{CN} \sum_\mu \sum_i
I(\vec{r_i^\mu}, \{ V_i^{\mu (k)} \}_{k = 1 \dots S})
\end{equation}
that is the information per synapse that can be extracted from $S$ different
observations of activity of the cells  with the animal is in $S$
different positions , and the system in activity states related 
to chart $\mu$.
This quantity does not diverge as $S \rightarrow \infty$, since
repeated observations of activity with the animal in nearby positions
do not yield independent information, because of correlations between
activity configurations, correlations which decrease with the distance
at 
which the
configurations are sampled.

The full calculation of this quantity involves a functional
integration over the  distribution of noise affecting cell activity as the
animal is moving, and exploring the whole environment. 
In Appendix \ref{app:info} we suggest a procedure to approximate this
quantity based on an ``information correlation length'' $l_I$ such
that samples corresponding to animal positions at a distance $l_I$
yield approximately independent information.

$I_s$ is the amount of spatial information which is stored
in the module. It is  the exact analogue of the stored information for
pattern auto-associators \cite{treves:cap90}. As for storage
capacity, it is to be found numerically, by maximization over $w$ and
$g$.

\begin{figure}[th]
\narrowtext
\begin{center}
\epsfig{file=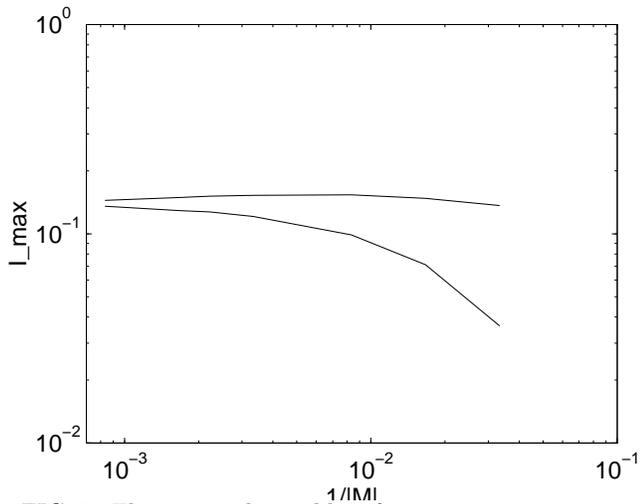, width=\columnwidth}
\caption[123]{The maximal storable information per synapse, as a
function of $1/|M|$.
} 
\label{fig:info1}
\end{center}
\end{figure}

As for the capacity one can find the solution which maximizes $I_s$.
The resulting $I_{\rm{max}}$ is a function of the size of the relative
spread of connections $a = 1 / |M|$, and it amounts to a fraction of
bit per synapse (see fig.~\ref{fig:info1}).

{\bf figure 5 approx. here}

As with pattern auto-associators, the information stored increases with
sparser representations. The increase is more marked for the fully
connected network. For very sparse representations the performance of
the fully connected model approaches the extreme dilution limit.

\section{Discussion}

We have studied the multi-chart threshold linear associator, as
a spatial information encoding and storage module. We have given the
solution for the dot-product kernel model, then we have introduced a
formalism in which the generic kernel problem is soluble. 

The second treatment has the advantage of
providing a form for the average activity peak profile, which can be
compared with the experimental data (see for example \cite{tus:sam97},
fig.1).

We have shown that the non-linear integral mean field equation (eq.~\ref{dilu:eqv})
can be solved at least for one class of connection kernels $K(r -
r')$.

The storage capacity for both models has been found. We note that the
capacity for the dot-product model is compatible with the wide
kernel (non sparse) limit of the generic model in one and two dimensions
in the fully connected and in the diluted condition.

The generic kernel treatment makes it possible to manipulate the most
relevant parameter for storage efficiency, i.e. the spread of
connections.
It is shown that this parameter plays a very similar role as
sparsity for pattern auto-associators. 
In the multi-chart case, moreover, the effective sparsity of the
{\em stable configurations} is determined also by the value of the gain
parameter $g$, as  shown analytically for the noiseless case.
Nevertheless, the capacity of the network depends on the spread of
connection parameter $a_c = k_d / |M|$ through a relation which is the
exact analogue of the relation between sparsity and capacity for
the pattern auto-associator, at least in the very sparse limit.

We have  only considered here the capacity problem
for one form of the connection kernel, although the treatment we
propose is applicable, at least, to the other kernels considered for the
noiseless case. 
Our  hypothesis is that a similar law for sparsity is to
be found as eq.~\ref{formalphac}, at least in the high sparsity limit,
for more general forms of the kernel.

We have then shown that the capacity scales in such a way that the information stored is not changed
when only a fraction of the cells participate in each chart.  In this case
the firing of a cell carries information not only about the position
of its p.f.c. in the chart environment, but also about {\em which }
environment the cell has a place-field in. This information adds up,
so that $1/a_c$ charts can be assembled in a single larger chart of
 size $1/a_c$ times larger.

We have introduced a definition of stored information for the multi-chart
memory network, which measures the number of effective different locations
which can be discriminated by such a net: representations of places at
a distance less than $l_I$ are confused, because of the finite width
 of the activity peaks, and because of the static noise.

$l_I$ does not vary much when $|M|$ varies.
This is consistent with the fact that the storage capacity is well
fitted by eq.~\ref{formalphac} with $k_d \sim 4.5$.
$l_I$ turns out to be $\sim 3.5$ for the 1-D model, with the arbitrary
value for $f$ of $0.95$.
$l_I$ is therefore similar to the ``radius'' of the activity peak
which should correspond to the ``pattern'' in the parallel between the
chart auto-associator and the pattern auto-associator.

It was not possible to carry over the calculation of $r_{12}$ and
$I_2$ in the 2-D model as it turns out to be too computationally
demanding. Therefore we are not able to show the values of the
storable information
The fact that the storage capacity follows eq.~\ref{formalphac} also
in this case is an indirect hint of a  behavior very similar to what
is found in 1-D.

\end{multicols}

\pagebreak

\begin{multicols}{2}

{\huge \bf \noindent Appendix}

\appendix

\section{Replica symmetric \\ free energy 
for the \\ ``dot product'' kernel model} \label{app:dotprod}

The replica symmetry free-energy reads

\beginwideapp
\begin{eqnarray}
f = &&-T\left\langle \left\langle \int Dz \ln {\rm Tr}(h,h_2)
\right\rangle \right\rangle - \frac{1}{2} \sum_{(\sigma),l}
|{\mathbf x}^{(\sigma),l}|^2 - B(x) - \sum_{(\sigma), l} (|{\mathbf s}^{(\sigma),l}| x +
{\mathbf s}^{(\sigma),l} x^{(\sigma), l}) - \sum_{(\sigma),l} {\mathbf t}^{(\sigma),l}
{\mathbf x}^{(\sigma), l} - tx \\ &&- r_0 y_0 + r_1 y_1 + \frac{\alpha d}{2
\beta} \left( \ln[1 - T_0 \beta (y_0 - y_1)] - \frac{ \beta y_1}{1 -
T_0 \beta (y_0 - y_1)} \right)
\end{eqnarray}
\endwideapp
very much like eq.19 in \cite{treves:cap90} and with the same meaning for
symbols, except that the population vector ${\mathbf x}^{(\sigma),l}$
plays the role of the overlap $x^\sigma$, the vector Lagrange
multiplier ${\mathbf t}^{(\sigma),l}$ appears instead of its scalar counterpart
$t^\sigma$ and the dimensionality $d$ appears multiplying the last
term.  $h$ and $h_2$ are
\begin{eqnarray}
h & = & -t - \sum_{(\sigma), l} {\mathbf t}^{(\sigma), l} \cdot
\vec{\eta}^{(\sigma), l} - z (-2 T r_1)^{1/2} \\ h_2 & = & r_1 -
r_0.
\end{eqnarray}
$\langle \langle \ldots\rangle \rangle$ means  averaging over
the distribution of p.f.c.'s $\vec{\eta}$.
$T$ is the noise level in the thermodynamic analysis.
$T_0$ is defined here as:
\begin{equation}
\left\langle \left\langle ({\mathbf t}_1^{l} \cdot
\vec{\eta}^{(\mu),l}) ({\mathbf t}_2^{l} \cdot
\vec{\eta}^{(\mu),l}) \right\rangle \right\rangle = \frac{T_0}{d}
{\mathbf t}_1^{l} \cdot {\mathbf t}_2^{l}
\end{equation}
and it is found to be equal to $1/2$ in 1-D and to $1$ for
the 2-D torus.

The saddle point equations can be found from this equations, and $t$
and ${\mathbf t}^{(\sigma),l}$ can be eliminated, in the same way as in
\cite{treves:cap90}.  Carrying on the calculation the $T=0$ equations eventually
reduce to two equations in the two variables (in the case of a single
``condensed'' map):

\begin{eqnarray}
w &=& \frac{[b(x) - \theta]}{ \rho} \\ 
{\mathbf v}^l &=&
\frac{({\mathbf x}^l + {\mathbf s}^l)}{ \rho}.
\end{eqnarray}

Take for simplicity $| {\mathbf v}^{l}| = v$ (while the direction is
set by ${\mathbf v}^{l}
\propto {\mathbf s}^{l}$).
The two equations read:
\begin{eqnarray}
&&E_1(w,{\mathbf v}) \equiv (A_1 + \delta A_2)^2 - \alpha A_3 = 0
\label{eqE1}\\ 
&&E_2(w,{\mathbf v})  \equiv \nonumber \\
&&(A_1 + \delta A_2) \left(
\frac{1}{g T_0 (1 + \delta)} + \alpha  - A_2 \right) - \alpha A_2 = 0 \label{eqE2}
\end{eqnarray}

where $\delta = |{\mathbf s}^1|/|{\mathbf x}^1|$ is the relative
importance of the external field and:
\beginwideapp
\begin{eqnarray}
A_1(w, v) & = & \frac{1}{v^2 T_0} \left\langle \left\langle
{\mathbf v}^{l} \cdot \vec{\eta}^{l} \int^+ Dz (w + \sum_l
{\mathbf v}^{l} \cdot \vec{\eta}^{l} -z) \right\rangle
\right\rangle - \left\langle \left\langle \int^+ Dz\right\rangle
\right\rangle \\ A_2(w, v) & = & \frac{1}{v^2 T_0} \left\langle
\left\langle {\mathbf v}^{(1)} \cdot \vec{\eta}^{(1)} \int^+ Dz (w +
\sum_l {\mathbf v}^{l} \cdot \vec{\eta}^{l} -z) \right\rangle
\right\rangle \\ A_3(w, v) & = & \left\langle \left\langle \int^+ Dz
(w + \sum_l {\mathbf v}^{l} \cdot \vec{\eta}^{l} -z)^2
\right\rangle \right\rangle
\end{eqnarray}  
$Dz$ is the Gaussian measure $(2 \pi)^{-1/2} e^{-z^2/2} dz$. The $+$
sign on the integral means that integration extremes are chosen such
that $(w + \sum_l {\mathbf v}^{l} \cdot \vec{\eta}^{l} -z) > 0$.

When the quenched average on the $\eta$'s is performed, $A_1$, $A_2$,
$A_3$ reduce to (for the $d$-dimensional torus ${\cal C}^d$):

\begin{eqnarray}
A_1(w, v) =&& \frac{1}{(2 \pi)^d v T_0} \int d\theta^{l} (\sum_l \cos
\theta^{l} )\times \nonumber \\ 
&&[ (w + v  \sum_l \cos \theta^{l} - v T_0) \Phi(w + v
\sum_l \cos \theta^{l} )  + (w + v  \sum_l \cos \theta^{l} )
\sigma(w + v  \sum_l \cos \theta^{l} )]
\end{eqnarray}
\begin{eqnarray}
A_2(w, v) =&& \frac{1}{(2 \pi)^d v T_0} \int d\theta^{l} (\sum_l \cos
\theta^{l} ) \times \nonumber \\
&& [ (w + v  \sum_l \cos \theta^{l} ) \Phi(w + v
\sum_l \cos \theta^{l} ) + (w + v  \sum_l \cos \theta^{l} )
\sigma(w + v  \sum_l \cos \theta^{l} )]
\end{eqnarray}
\begin{eqnarray}
A_3(w, v) =&& \frac{1}{(2 \pi)^d} \int d\theta^{l} \times \nonumber \\
 &&[1 + (w + v  \sum_l
\cos \theta^{l} )^2] \Phi(w + v  \sum_l \cos \theta^{l} ) +
(w + v  \sum_l \cos \theta^{l} )\sigma(w + v  \sum_l \cos \theta^{l} ) 
\end{eqnarray}
\endwideapp
\noindent where

\begin{eqnarray}
\label{eqPhi} \Phi(x) &=& \int_{-\infty}^{x} \frac{dz}{\sqrt{2\pi}}
e^{\frac{-z^2}{2}}   \\
\sigma(x) &=& \frac{e^{\frac{-x^2}{2}}}{\sqrt{2\pi}}. \label{eqsigma}
\end{eqnarray}

\section{Generic kernel, extreme dilution} \label{app:gendilu}

Let us consider the one-dimensional case first, and consider the
kernel
\begin{equation}
K(\vec{r} - \vec{r}')  = \hat{K}(\vec{r} -
\vec{r}') - \frac{2}{|M|} = e^{-| \vec{r} -
\vec{r}' |} - \frac{2}{|M|}.
\end{equation}

Eq.~\ref{dilu:eqv} can be written
\begin{equation}
v(\vec{r}) = g {\cal N}\left( \int d \vec{r}' \hat{K}(\vec{r} -
\vec{r}') v(\vec{r}') + \hat{w} \right) \label{dilu:eqhat}
\end{equation}
where
\begin{equation}
\hat{w} = w - \frac{2}{|M|} \int d \vec{r}' v(\vec{r}'). \label{dilu:eqw}
\end{equation}
For the purpose of finding $\alpha_c$, maximizing with respect to
$\hat{w}$ is equivalent to maximizing with respect to $w$.

To solve eq.\ref{dilu:eqv}, the transformation
\begin{equation}
u(\vec{r}) = {\cal N}^{-1} \left(\frac{v(\vec{r})}{g}
\right) \label{dilu:transfu}
\end{equation}
is used, which results in
\begin{equation}
u(\vec{r}) = g \int d \vec{r}'  \hat{K}(\vec{r} -
\vec{r}') {\cal N}[u(\vec{r})] + \hat{w}  \label{dilu:equ}.
\end{equation}
By differentiating twice we get
\begin{equation}
u''(\vec{r}) = -2 g {\cal N} [u(\vec{r})] +
u(\vec{r}) - \hat{w} =  - \frac{d}{du} {\cal U}[u(\vec{r})] \label{dilu:diffequ}
\end{equation}
where
\begin{equation}
{\cal U} = \int^u du' \left( 2g {\cal N} (u') - u' + \hat{w} \right). \label{eq:potential}
\end{equation} 

The differential equation (\ref{dilu:diffequ}) is {\em locally}
equivalent to the non-linear integral equation (\ref{dilu:equ})
Equation (\ref{dilu:diffequ}) must be solved numerically. As in the single map case,
not all the solutions of the differential equation
(\ref{dilu:diffequ}) are solution of the integral equation
(\ref{dilu:equ}).
Solutions of (\ref{dilu:diffequ}) are solution of (\ref{dilu:equ}),
strictly speaking, only in the case $M \equiv \Bbb{R}^d$.
Nevertheless, we force the equivalence since,   also in the case of limited
environments, with periodic boundary conditions, 
possible pathologies are not important for solutions with activity
concentrated  far from the boundaries.

In order to classify the
solutions of eq.~\ref{dilu:diffequ} it is useful to study the
``potential function'' ${\cal U}$.  If $w$ is negative and large
enough in absolute value, ${\cal U}(u)$ has a maximum and a minimum at
the two roots of equation
\begin{equation}
\frac{d}{du} {\cal U}(u) = 2 g {\cal N} (u) - u + \hat{w} = 0,
\end{equation}
or, in terms of $v$:
\begin{equation}
v = g{\cal N}(2v + \hat{w}), \label{dilu:eqconstv}
\end{equation}
corresponding to constant solutions of eq.~\ref{dilu:eqv}.  We look
for solutions representing a single, symmetric peak of activity
centered in $r = 0$. We therefore need to solve the Cauchy problem
given by eq.~\ref{dilu:diffequ} with the initial conditions:
\begin{eqnarray} 
u(0) &=& u_0 \\ u'(0) &=& 0
\end{eqnarray}

\begin{figure}[ht]
\narrowtext
\begin{center}
\epsfig{file=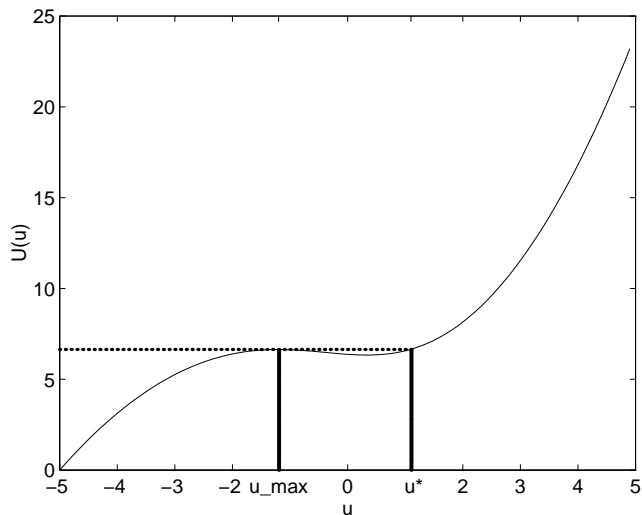, width = \columnwidth}
\caption{The ``potential'' function ${\cal U}(u)$ defined by
eq.~\ref{eq:potential} and entering the differential equation 
eq.~\ref{dilu:diffequ}. Solutions with $u'(0) = 0$ and $u(0) = u_0$,
with $u_{max} < u_0 < u^*$ are oscillating. The solution with $u_0 =
u^*$ is the one we seek, asymptotically approaching $u_{max}$ as $r
\rightarrow \infty$.} \label{fig:potential}
\end{center}
\end{figure}

From fig.~\ref{fig:potential} it is clear that if $u_0 > u^*$ the solution
will escape to $- \infty$ for $r$ tending to infinity.  This will
correspond to $v$ tending asymptotically to 0, and this solution cannot be a
solution for the integral equation (\ref{dilu:eqv}) as the asymptotic
value must be a root of (\ref{dilu:eqconstv}).

The solutions of the problem with $u_0 < u^*$ are periodic,
corresponding to multiple peaks of activity, and they are discarded as
unstable with the same arguments holding for the single map case.
There is also the constant solution 
\begin{equation}
u(r) = u_{min},
\end{equation}
which obviously will not correspond to space related activity.
The solution corresponding to the single activity peak can only be the
one with $u_0 = u^*$.
It tends asymptotically to $u_{max}$.
This solution can be found numerically and inserted in eq.\ref{dilu:eqalpha} to
find the value of $\alpha$ associated with the pair $(g, \hat{w})$.
The solution will only be present for values  of $\hat{w}$ for which ${\cal
U}(u)$ has the extremal
points $u_{max}$ and $u_{min}$, that is:
\begin{equation} 
\hat{w} < \hat{w}^*
\end{equation}
where $w^*$ is equal to $-2 g {\cal N}(u^*) + u^*$ and $u^*$ is the root of the equation:
\begin{equation}
\Phi(u) = \frac{1}{2g}
\end{equation}
obtained by derivating twice ${\cal U}$,
and this shows that eq.~\ref{dilu:equ} cannot have solutions for $g <
1/2$, as in the single map case.

In the two dimensional case, we can consider the kernel
\begin{equation}
K(\vec{r} - \vec{r}') = \hat{K}(\vec{r} -
\vec{r}') - \frac{2}{|M|}
\end{equation}
where $\hat{K}$ is the kernel having Fourier transform:
\begin{equation}
\hat{K}({\mathbf p}) = \frac{2}{1 + {\mathbf p}^2}
\end{equation}
The solution is worked out in the same way with the transformation
(\ref{dilu:transfu}) and application of Laplacian.
If we consider solutions with circular symmetry and pass to polar
coordinates $(r, \phi)$, the equation for $r$ reads:
\begin{equation}
u''(r) + \frac{1}{r} u'(r) = -2 g {\cal N} [u(r)] +
u(r) - w
\end{equation}
We still have a single peak solution with tends asymptotically to
$u_{max}$, but in this case we cannot rely on the ${\cal U}$ function
argument to find the initial condition at $r = 0$, which has to be
found numerically.

{\bf Figure 6 approx. here}

\section{Replica free energy calculation for the generic kernel} \label{app:genfull}

Again we will consider an environment $M$ with periodic boundary conditions.
We assume that there exists a kernel $L$ such that
\begin{equation}
\int d \vec{r}'' L(\vec{r} - \vec{r}'')
L(\vec{r}'' - \vec{r}') = |M| K(\vec{r} - \vec{r}').
\end{equation} 

Instead of the vector order parameter ${\mathbf x}^\mu$ we used for the
dot-product kernel case (or of the scalar overlap $x^\mu$ of
\cite{treves:cap90})
we can use the functional order parameter 
\begin{equation}
x^\mu (\vec{r}) = \frac{1}{N} \sum_i [L(\vec{r} - \vec{r}^\mu_i)] V_i
\end{equation}
in terms of which the interaction part of the Hamiltonian (\ref{fc:ham})
reads
\beginwideapp
\begin{eqnarray}
&& \frac{1}{2} \sum_i \sum_{j \neq i} J_{ij} V_i V_j = \nonumber \\
&&\frac{|M|}{2 N} \sum_\mu  \sum_i \sum_j \left[
K(\vec{r}^\mu_i - \vec{r}^\mu_j) - \bar{K} \right] V_i V_j
- \frac{\alpha |M|}{2 } (K(0) - \bar{K}) \sum_i V_i^2 = \nonumber \\
&& \frac{1}{2} N \sum_\mu \int d \vec{r} \left[ x^\mu (
\vec{r} ) \right]^2 - \frac{\alpha |M|}{2} (K(0) - \bar{K})
\sum_i V_i^2
\end{eqnarray}
Introducing the ``square root'' kernel $L$ allows us to perform the standard Gaussian
transformation manipulation and to carry out the mean field free
energy calculation in the replica symmetry approximation:
\begin{eqnarray}
f =&& -T\left\langle \left\langle \int Dz \ln {\rm Tr}(h,h_2)
\right\rangle \right\rangle - \frac{1}{2} \sum_{\sigma}
\int_M d \vec{r} \left[ x^\sigma (\vec{r}) \right]^2 -
\frac{\alpha |M| }{2} (K(0) - \bar{K}) y_0 + B(x) - \nonumber \\
&&\sum_{\sigma} \int_M d \vec{r} \; t^{\sigma} (\vec{r} )
x^{\sigma} (\vec{r} ) - tx  - r_0 y_0 + r_1 y_1 + \nonumber \\ 
&&\frac{\alpha }{2\beta}
 \sum_{{\mathbf p}} \left( \ln[1 - T_0({\mathbf p}) \beta (y_0 - y_1)] - \frac{ \beta y_1}{1 -
T_0({\mathbf p}) \beta (y_0 - y_1)} \right)
\end{eqnarray}
where now $T_0({\mathbf p})$ is the Fourier transform of the kernel
$|M| \hat{K}$
\begin{equation}
T_0(p) = |M| \int_M d \vec{r} e^{-i {\mathbf p}
\vec{r} } K(\vec{r}).
\end{equation}

We now have
\begin{eqnarray}
h &=& b(x) + \sum_\sigma \int_M d \vec{r}' x^\sigma (
\vec{r}') \left[ L(\vec{r}^\sigma - \vec{r}') -
\bar{L} \right] - z (-2 t r_1)^{1/2} \\
h_2 &=& -r_0+ r_1.
\end{eqnarray}

The $T=0$ mean field equations are much like in \cite{treves:cap90} apart
from the $x^\sigma(\vec{r})$ equation which reads:
\begin{equation}
x^\sigma(\vec{r}) = g' \left\langle \left\langle \left[ L(\vec{r}^\sigma - \vec{r}') -
\bar{L} \right] \int^+ Dz \left\{ \int_M d
\vec{r}' \left[ L(\vec{r}^\sigma - \vec{r}') -
\bar{L} \right] x^\sigma(\vec{r}') +b(x) - \theta - \rho z \right\}
\right\rangle \right\rangle
\end{equation}
where now the $+$ sign on the integral means that the limits of
integration over $z$ are chosen such that
\begin{equation}
\int d
\vec{r}' \left[ L(\vec{r}^\sigma - \vec{r}') -
\bar{L} \right] x^\sigma(\vec{r}') +b(x) - \theta > 0.
\end{equation}
$g'$ is a renormalized gain, which takes into account the effect of
static noise, defined by:
\begin{equation}
(g')^{-1} = g^{-1} - \alpha \sum_p T_0(p) \frac{\bar{\Psi}}{1 - T_0(p)\bar{\Psi}} 
\end{equation}
where $\bar{\Psi}$ is given by eq.~\ref{fc:eqpsi}.

The noise variance $\rho^2$ is given by 
\begin{equation}
\rho^2 = - 2 T r_1 =  \alpha \sum_{{\mathbf p}} \frac{[T_0(p)]^2
y_0}{\left[ 1
- T_0(p) \bar{\Psi} \right]^2} \label{fc:eqrho}
\end{equation}
where 
\begin{equation}
y_0 = (g')^2 \left\langle \left\langle  \int^+ Dz \left\{ \int_M d
\vec{r}' \left[ L(\vec{r}^\sigma - \vec{r}') -
\bar{L} \right] x^\sigma(\vec{r}') +b(x) - \theta \right\}^2 \right\rangle \right\rangle
\end{equation}
and 
\begin{equation}
\bar{\Psi} = g' \left\langle \left\langle  \int^+ Dz 
\right\rangle \right\rangle.
\end{equation}

We now pass to the rescaled variables 
\begin{eqnarray}
v^\sigma (\vec{r}) &=& \frac{x^\sigma(\vec{r})}{\rho}\\
w &=& \frac{b(x)}{\rho}
\end{eqnarray}
obtaining
\begin{eqnarray}
v^\sigma (\vec{r}) &=& g' \int_M d \vec{r}^\sigma \left[
L(\vec{r}^\sigma - \vec{r}) - \bar{L} \right]
{\cal N} \left( w + \int_M d \vec{r}' \left[
L(\vec{r}^\sigma - \vec{r}') - \bar{L} \right]
v^\sigma(\vec{r}) \right) \label{fc:eqv} \\
\frac{y_0}{\rho^2} &=& {(g')^2} \int_M \frac{d \vec{r}^{\sigma}}{|M|}  {\cal M} \left( w + \int_M d \vec{r}' \left[
L(\vec{r}^\sigma - \vec{r}') - \bar{L} \right]
v^\sigma(\vec{r}) \right) \label{fc:eqy} \\
\bar{\Psi} &=&  \int_M \frac{d \vec{r}^{\sigma}}{|M|} \Phi\left( w + \int_M d \vec{r}' \left[
L(\vec{r}^\sigma - \vec{r}') - \bar{L}  \right] \right)
v^\sigma(\vec{r}). \label{fc:eqpsi}
\end{eqnarray}

\endwidebottomapp

\section{Generic kernel: \\ storable information calculation} \label{app:info}

First, the information per synapse we get from a single observation
of activity,
with the animal in a certain position times the number of stored charts is 

\begin{eqnarray}
&&I_1 = \nonumber \\
&& \alpha \int \frac{d\vec{r}}{ |M|} \left\{
   \int_{-\infty}^{u(\vec{r})} \frac{dz}{ \sqrt{2\pi}}e^{-\frac{z^2}{2}} \log \left( 
                          \frac{e^{-\frac{z^2}{2}}}
                          {\int {\frac{d\vec{r'}}{
|M|}}{e^{-\frac{(z-u(\vec{r}) +u(\vec{r'})^2)}{2}} } } \right)
\right. \nonumber\\
&&  + \left . \left[ 1-\phi(u(\vec{r}))\right] \log \left(
                          \frac{\left[ 1-\phi(u(\vec{r}))\right] }{          
                          \int {\frac{d\vec{r'}}{ M}}\left[ 1-\phi(u(\vec{r'})) \right]} \right) \right\}.
\end{eqnarray}

Next, we wish to calculate the joint information from two measures of
activity, from the same cells, from all charts, while the rat is in two different
locations, at a distance $\epsilon$.
These two measures are  correlated random variables: let 
\[
V_1 = [h_1 - \rho z_1]^+
\]
be the activity of a cell  measured while the rat is in position $1$, and 
\[
V_2 = [h_2 - \rho z_2]^+
\]
be the activity of the same cell while the rat is in position $2$.

The two noise variables are distributed according to a joint bivariate
gaussian distribution:
\begin{eqnarray}
&&p(z_1, z_2) = \nonumber \\
&& \frac{1}{2 \pi \sqrt{1 - r_{12}^2}} \exp\left(-
\frac{1}{2 (1 - r_{12}^2)} (z_1^2 + z_2^2 - 2 r_{12} z_1 z_2)
\right).
\end{eqnarray}

The correlation coefficient $r_{12}$ is a function of the distance
$\epsilon$, implicitly defined through the equation

\begin{equation}
\rho^2 r_{12}(\epsilon) = \alpha |M| \langle \langle K^2  \rangle \rangle
y_{12}(\epsilon)  \label{info:eqrho1}
\end{equation}
where $y_{12}$is defined as 

\begin{equation}
y_{12}(\epsilon) = \frac{1}{N} \sum_i \langle V_{i,1} V_{i,2} \rangle 
\end{equation}
and assuming periodic boundary conditions:
\beginwideapp
\begin{eqnarray}
y_{12}(\epsilon) =&& \rho^2 g^2 \int \frac{d r}{|M|} \int^{++}
Dz_{12} \times \\
&&\left( \int \frac{d r'}{|M|} K(r - r') v(r') + w - z_1
\right) 
 \left( \int \frac{d r''}{|M|} K(r - r'') v(r''+\epsilon) + w - z_2
\right) 
\end{eqnarray}
\endwideapp
or,
\begin{equation}
y_{12}(\epsilon) = \rho^2 g^2 \int \frac{d r}{|M|} \int^{++}
Dz_{12} u(r) u(r + \epsilon) \label{info:eqy12}
\end{equation}
where $u(r)$ is defined by eq.~\ref{dilu:transfu}.
The integration measure  for the noise variable is defined as 
\begin{equation}
\int^{++} Dz_{12} = \int_{u(r) - z_1 > 0, u(r + \epsilon) - z_2
> 0} dz_1 dz_2 p(z_1, z_2). \label{info:zmeas}
\end{equation}

Inserting eq.~\ref{info:eqy12} in eq.~\ref{info:eqrho1} we yield:
\begin{equation}
r_{12} = \alpha |M| \langle \langle K^2 \rangle \rangle g^2
\int d r {\cal Q} (u(r), u(r + \epsilon), r_{12}) \label{info:eqrho}
\end{equation}
where 
\[
{\cal Q} (x, y, r_{12}) = \int^{++} Dz_{12} (x - z_1)(y - z_2).
\]
 
\begin{figure}[ht]
\narrowtext
\begin{center}
\epsfig{file=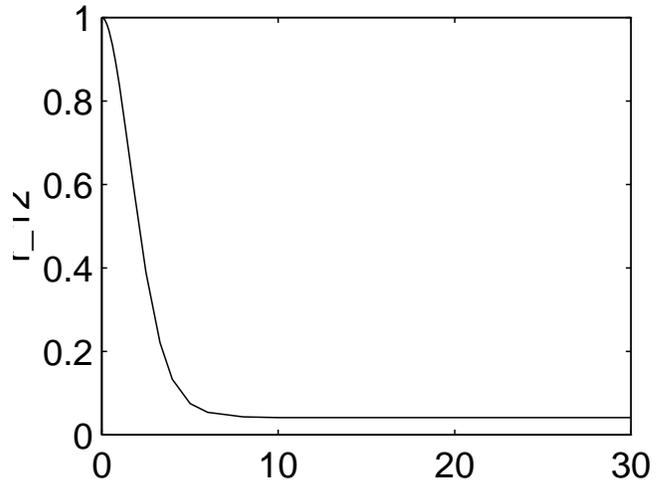, width = \columnwidth}
\caption{The $r_{12}$ function plotted as a function of the distance
between the two p.f.c.s, $\epsilon$ in the $|M| = 30$ case.} \label{fig:rho12}
\end{center}
\end{figure}

Eq.~\ref{info:eqrho} can be solved numerically, an example is provided
in fig.~\ref{fig:rho12}, but a few features can be explored analytically, in the
neighborhood of $\epsilon = 0$.
$r_{12} = 1, \epsilon = 0$ is a solution, but now consider what
happens when $\epsilon$ increases.

{\bf Figure 7 approx. here}

The derivatives of 

\begin{equation}
{\cal D}(r_{12}, \epsilon) =  \alpha \langle \langle K^2 \rangle \rangle g^2
\int \frac{d r}{|M|} {\cal Q} (u(r), u(r + \epsilon), r_{12})
- r_{12}
\end{equation}
with respect to $\epsilon$ and $r_{12}$ must be taken into
consideration.
One has:
\begin{eqnarray}
&& \frac{\partial}{\partial \epsilon} {\cal D} (r_{12} = 1,\epsilon = 0) = 
\alpha \langle \langle K^2 \rangle \rangle g^2
 \int \frac{d r }{|M|} \times \nonumber \\
&& \frac{\partial}{\partial y} {\cal Q}(x = u(r), y = u(r +
 \epsilon), 1) u'(r) = 0,
\end{eqnarray}

\begin{equation}
 \frac{\partial^2}{\partial \epsilon^2} {\cal D} (r_{12} =
1,\epsilon = 0) < 0 \label{eq:d2eps}
\end{equation}
and 
\begin{eqnarray}
&&\frac{\partial}{\partial r_{12}} {\cal D} (r_12 \rightarrow 1,
\epsilon = 0) = \nonumber \\
&&\alpha \langle \langle K^2 \rangle \rangle g^2
 \int \frac{d r }{|M|} \Phi(u(r)) - 1 = \nonumber \\
&&\int \frac{d r }{|M|} \Phi(u(r))  \left(\int \frac{d r }{|M|}
 {\cal M}(u(r))\right)^{-1} - 1 \label{eq:ddrho}
\end{eqnarray}
From eq.~\ref{eq:d2eps} it turns out that when the derivative in
eq.~\ref{eq:ddrho} is greater than zero, the solution $r_{12} = 1$
disappears as one moves from $\epsilon = 1$, but another solution is
still present so that 
\begin{equation}
\lim_{\epsilon \rightarrow 0^+} r_{12} (\epsilon) < 1. \label{eq:limrho12}
\end{equation}

Note that the condition 
\[
 \frac{\partial}{\partial r_{12}} {\cal D} (r_{12} \rightarrow 1,
\epsilon = 0) > 0
\]
is equivalent to 
\begin{equation}
\Gamma = g \alpha |M| \langle \langle K^2 \rangle \rangle \int_M dr\,
u(r)v(r) < 0,
\end{equation}
and the quantity $\Gamma$ enters in the stability analysis
considerations we sketched in sec.~\ref{sec:stability}, at least for
the 1-D case. Solutions with
$\Gamma > 0$ are  stable against inhibition
orthogonal fluctuations, so that it is likely that the possible pathology
implied by eq.~\ref{eq:limrho12} reflects an instability of the solution.
We have always found numerically  that for solution corresponding to the maximal storage
capacity and information, $\Gamma > 0$.

Once we know the joint probability distribution for $z_1$ and $z_2$, we
can calculate the information we can extract about the p.f.c. of a
cell from two measurements of activity, while the rat is standing in
two positions at a distance $\epsilon$, from all charts.

\beginwidetopapp
\begin{eqnarray}
I_2(\epsilon) =&& \alpha \int_M \frac{d r}{|M|} \Biggl\{ \int^{++} Dz_{12}  
 \Bigl[ -\frac{1}{2 (1 - r_{12}^2)} (z_1^2 + z_2^2 - 2
r_{12} z_1 z_2) - \nonumber \\
&& \log  \Bigl( \int_M  \frac{d r'}{|M|} \exp \bigl( - \frac{1}{2 (1 -
r_{12}^2)} [(u(r') - u(r) + z_1)^2 + (u(r' + \epsilon) -
u(r + \epsilon) + z_2)^2 \nonumber    \\
  &&- 2 r_{12} (u(r') - u(r) + z_1)
(u(r' + \epsilon) - u(r + \epsilon) + z_2) ]  \bigr)
\Bigr) \Bigr]  \nonumber \\
&&+ 2  \int^{+-} Dz_{12} \biggl[ \log \Bigl( \int_{z_2' > u(r +
\epsilon)} \frac{d z_2'}{2 \pi} \exp \bigl(-
\frac{1}{2 (1 - r_{12}^2)} (z_1^2 + {z_2'}^2 - 2 r_{12} z_1 z_2')
\bigr) \Bigr) \nonumber \\
&&- \log \Bigl( \int_M  \frac{d r'}{|M|} \int_{z_2' > u(r' +
\epsilon)} d z_2' \exp \bigl( - \frac{1}{2 (1 -
r_{12}^2)} [(u(r') - u(r) + z_1)^2 + (u(r' + \epsilon) -
u(r + \epsilon) + z_2')^2   \nonumber  \\
&&- 2 r_{12} (u(r') - u(r) + z_1)
(u(r' + \epsilon) - u(r + \epsilon) + z_2') ] \bigr) \Bigr)
\biggr] \nonumber \\
&&+ \int^{--} Dz_{12} \biggl[ \log \Bigl( \int_{z_1' >
u(r),z_2' > u(r + \epsilon)} \frac{dz_1' dz_2'}{2 \pi \sqrt{1 -
r_{12}^2} }     \exp \bigl(-
\frac{1}{2 (1 - r_{12}^2)} ({z_1'}^2 + {z_2'}^2 - 2 r_{12} z_1' z_2')
\bigr) \Bigr) \nonumber \\
&&- \log \Bigl( \int_M \frac{d r'}{|M|} \int_{z_1' >
u(r'),z_2' > u(r' + \epsilon)}\frac{dz_1' dz_2'}{2 \pi \sqrt{1 -
r_{12}^2} } \exp \bigl( - \frac{1}{2 (1 -
r_{12}^2)}   \nonumber \\
 &&[(u(r') - u(r) + z_1')^2 + (u(r' + \epsilon) -
u(r + \epsilon) + z_2')^2    \nonumber  \\
   &&- 2 r_{12} (u(r') - u(r) + z_1')
(u(r' + \epsilon) - u(r + \epsilon) + z_2') ] \bigr)
\Bigr) \biggr] \Biggr\}.
\end{eqnarray}
\endwideapp

The minus signs ($-$) beside the integration signs mean that 
respectively the first, or the second condition determining the
integration intervals in eq.~\ref{info:zmeas} are reversed. The first
term in the sum accounts for the contribution coming from 
measurement in which both activity values are positive. The second
term is the contribution from measurements in which one value is zero and 
the other is positive.
The third term comes from measurements in which both values are zero.

For $\epsilon = 0$, $I_2 = I_1$, since the two measures are
identical.

\begin{figure}[ht]
\narrowtext
\begin{center}
\epsfig{file=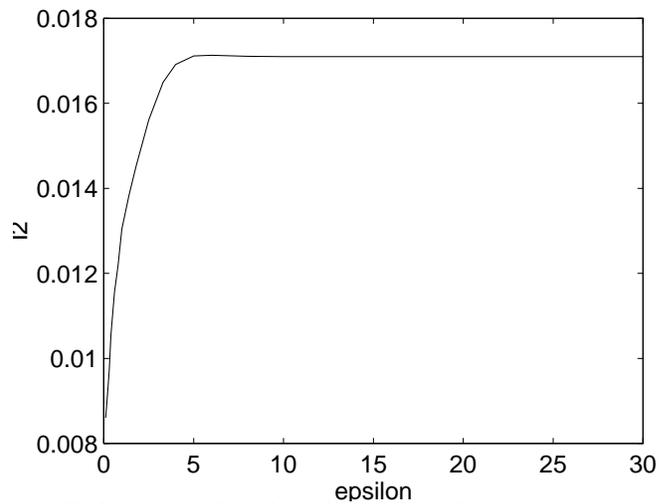, width = \columnwidth}
\caption{The $I_2$ function plotted as a function of the distance
between the two p.f.c.s, $\epsilon$ in the $|M| = 30$ case.
Note that $l_I$, with $f = 0.95$ (see eq.~\ref{app:eqli}) would be
approximately 3.5. This is seen not to change much when $|M|$ is
varying (not shown). } \label{fig:I2}
\end{center}
\end{figure}

For large $\epsilon$ one has $I_2 \sim 2 I_1$, because the noise
decorrelates and because in general the two measures will give
non-zero results in distinct regions of the environment. 
The behavior of $I_2$ as a function of $\epsilon$ is exemplified in
figure \ref{fig:I2}.
We define as ``information correlation length'' the value 
$l_I$ of  $\epsilon$ for which
\begin{equation}
I_2 - I_1 = f I_1, \label{app:eqli}
\end{equation}
where $f$ is a fixed fraction, say 0.95.
Note that this quantity 
We may say that measurements of activity with the rat in two positions
at a distance $l_I$ give independent information.

{\bf Figure 8 approx. here}

This allows  us to define as the {\em stored information} $I_s$ the
quantity
\begin{equation}
I_s = I_1 \frac{|M|}{l_I^d},
\end{equation} 
that is, sampling the activity of a cell $\frac{|M|}{l_I^d}$ times,
with the animal spanning a lattice with size $l_I$, we may effectively
add up the information amounts we get from each single sample, as if
they were independent.

\bibliographystyle{prsty}

\bibliography{med1,med2,med3,ann,georgo}

\begin{thebibliography}{10}

\bibitem{amit:book}
D. Amit, {\em Modeling Brain Function.} (Cambridge University Press, New York,
  1989).

\bibitem{tso:elba94}
M. Tsodyks and T. Sejnowski, Proceedings of the third workshop on Neural
  Networks: from Biology to High Energy Physics - International Journal of
  Neural Systems {\bf 6 Supp.},  81  (1994).

\bibitem{taube:hdcrev96}
R. Muller, J. Ranck, and J. Taube, Current Opinion in Neurobiology {\bf 6},
  196  (1996).

\bibitem{redish:navig96}
D. Touretzky and A. Redish, Hippocampus {\bf 6},  247  (1996).

\bibitem{tus:got97}
K. Gothard, W. Skaggs, and B. McNaughton, Journal of Neuroscience {\bf 16},
  8027  (1996).

\bibitem{treves:cap90}
A. Treves, Physical Review A {\bf 42},  2418  (1990).

\bibitem{derrida:dilu87}
B. Derrida, E. Gardner, and A. Zippelius, Europhysics Letters {\bf 4},  167
  (1987).

\bibitem{treves:dilu91}
A. Treves, Journal of Physics A: Math. Gen. {\bf 24},  327  (1991).

\bibitem{tus:sam97}
A. Samsonovich and B. McNaughton, Journal of Neuroscience {\bf 17},  5900
  (1997).

\bibitem{geopopvec}
A. Georgopoulos, R. Kettner, and A. Schwartz, Journal of Neuroscience {\bf 8},
  2928  (1988).

\bibitem{wilmn93}
M. Wilson and B. McNaughton, Science {\bf 261},  1055  (1993).

\end{thebibliography}

\pagebreak

\end{multicols}

\end{document}